\documentclass[aps, prb, twocolumn, amssymb, amsmath, superscriptaddress]{revtex4-1}

\usepackage{bm}
\usepackage{times}
\usepackage{graphicx}
\usepackage[colorlinks,citecolor=blue,linkcolor=blue]{hyperref}
\usepackage{array}
\usepackage{float}
\usepackage{colortbl}
\usepackage{multirow}
\usepackage{tabularx}

\begin{document}

\title{Electronic Transport of Two-Dimensional Ultrawide Bandgap Material h-BeO}

\author{Yanfeng Ge}
\affiliation{State Key Laboratory of Metastable Materials Science and Technology \& Key Laboratory for Microstructural Material Physics of Hebei Province, School of Science, Yanshan University, Qinhuangdao, 066004, China}

\author{Wenhui Wan}
\affiliation{State Key Laboratory of Metastable Materials Science and Technology \& Key Laboratory for Microstructural Material Physics of Hebei Province, School of Science, Yanshan University, Qinhuangdao, 066004, China}

\author{Yulu Ren}
\affiliation{State Key Laboratory of Metastable Materials Science and Technology \& Key Laboratory for Microstructural Material Physics of Hebei Province, School of Science, Yanshan University, Qinhuangdao, 066004, China}

\author{Fei Li}
\affiliation{State Key Laboratory of Metastable Materials Science and Technology \& Key Laboratory for Microstructural Material Physics of Hebei Province, School of Science, Yanshan University, Qinhuangdao, 066004, China}

\author{Yong Liu}\email{yongliu@ysu.edu.cn}
\affiliation{State Key Laboratory of Metastable Materials Science and Technology \& Key Laboratory for Microstructural Material Physics of Hebei Province, School of Science, Yanshan University, Qinhuangdao, 066004, China}

\date{\today}

\begin{abstract}
Two-dimensional ultrawide bandgap materials, with bandgaps significantly wider than 3.4 eV, have compelling potential advantages in nano high-power semiconductor, deep-ultraviolet optoelectronics, and so on. Recently, two-dimensional layered h-BeO has been synthesized in the experiments. In the present work, the first-principles calculations predict that monolayer h-BeO has an indirect bandgap of 7.05 eV with the HSE functional. The ultrawide bandgap results from the two atomic electronegativity difference in the polar h-BeO. And the electronic transport properties are also systematically investigated by using the Boltzmann transport theory. The polar LO phonons of h-BeO can generate the macroscopic polarization field and strongly couple to electrons by the Fr\"{o}hlich interaction. Limited by the electron-phonon scattering, monolayer h-BeO has a high mobility of 473 cm$^2$V$^{-1}$s$^{-1}$ at room temperature. Further studies indicate that the biaxial tensile strain can reduce the electronic effective mass and enhance the electron-phonon coupling strength. The suitable strain can promote the mobility to $\sim$1000 cm$^2$V$^{-1}$s$^{-1}$ at room temperature.
\end{abstract}

\maketitle

Since the emergence of the germanium-based transistor at Bell Telephone laboratories, modern semiconductor technologies span a huge range of applications due to their manipulation of electrons, holes, and photons in a wide variety of device architectures and operating environments. With the fast development of electronics industries, high-performance power semiconductor devices become particularly important. As an important criterion to contrast candidate materials for the power-electronic devices, Baliga's figure of merit depends most strongly on the breakdown field, which increases nonlinearly with increasing bandgap. So there is a huge demand for ultrawide bandgap (UWBG) semiconductors~\cite{Tsao2018}, those have bandgaps significantly wider than the 3.4 eV of GaN. Such as, AlGaN/AlN, diamond, Ga$_2$O$_3$, BN and II-IV-N materials~\cite{Xue2018,Wang2020,Chae2019,Chikoidze2019}. Also with further exploration, the UWBG semiconductors have significant application prospects in radio frequency and microwave electronics, deep-ultraviolet optoelectronics, quantum information, and extreme-environment applications~\cite{Kim2017}.
Among the UWBG materials, BN is of special interest. Its wurtzite (w-BN), zinc blende (c-BN), and hexagonal (h-BN) all belong to UWBG semiconductors. The h-BN has strong in-plane covalent bonds and weak Van der Waals force between layers~\cite{Zhang2017,Wang2019,Wang2017,Song2010}. The similar structure to graphene also makes h-BN be well-known as 'white graphene'. And it especially attracts for 2D materials because of the unique physical and chemical properties~\cite{Liu2020,Topsakal2009,Ares2020}, such as the atomic flatness, mechanical robustness, absence of dangling bonds, and high thermal conductivity. So of course, in the recently rapidly developing field of layer engineering of 2D semiconductor conjunctions, there is a large amount of h-BN-based van der Waals heterostructures~\cite{Zhao2018,Vu2016,Wang2015,Cheng2017}, which greatly expands its application fields.

The h-BN has a wide range of potential applications and also provides strong motivation for the exploration of novel 2D UWBG materials. Inspired by the element changes from graphene to h-BN and the bandgap in h-BN, it comes naturally that the compound of beryllium and oxygen is very likely another 'white graphene'. The early research work finds that the sp$^3$-hybridized w-BeO is also insulator with high thermal conductivity and the sp$^2$-hybridized h-BeO is synthetized hardly in spite of the predictive UWBG~\cite{Continenza1990,Freeman2006}. Until recently, Wang~\cite{LWang2020} obtains the layered h-BeO in the graphene-encapsulated confined cell, which produces a miniaturized high-pressure container for the crystallization in solution. And the energy barrier between the w- and h-BeO phases is responsible for the observed h-BeO layer counts beyond the ultra-thin limit in the theoretical prediction. Other theme that come up, many physical properties of a newly synthesized material are not implemented yet and looking forward to being realized in the future. The comprehensive estimation and understanding of carrier transport are essential for the application potential in multifunctional electronic devices, especially power electronics.

In this work, we have studied the electronic structure, phonon dispersions, electron-phonon coupling, and carrier mobility of monolayer h-BeO using first-principles calculations with Boltzmann transport theory. The present results show that monolayer h-BeO is a UWBG material with an indirect bandgap of 7.05 eV in the HSE functional, which results from the electronegativity difference between Be and O atoms. Equally important is that the polar LO phonons in monolayer h-BeO can couple with the electron strongly by the Fr\"{o}hlich interaction. On this basis, monolayer h-BeO has a high mobility of 473 cm$^2$V$^{-1}$s$^{-1}$ at room temperature. More interesting, a suitable biaxial tensile strain can promote the mobility to $\sim$1000 cm$^2$V$^{-1}$s$^{-1}$.

All calculations were carried out in the framework of density-functional theory (DFT) as implemented in the QUANTUM ESPRESSO package~\cite{Giannozzi2017}. The exchange and correlation energy was in the Perdew-Burke-Ernzerhof (PBE) form~\cite{Perdew1996}. And the more accurate value of the bandgap was obtained by using hybrid Hartree-Fock+DFT functional of HSE~\cite{Heyd2003}. By requiring the convergence of mobility, the kinetic energy cutoff of 40 Ry, 320$\times$320$\times$1 k-mesh, and 80$\times$80$\times$1 q-grid were used in EPW code~\cite{Ponce2016}. The other computational details are given in the supplementary material.

\begin{figure}[htp!]
\includegraphics[width=0.5\textwidth]{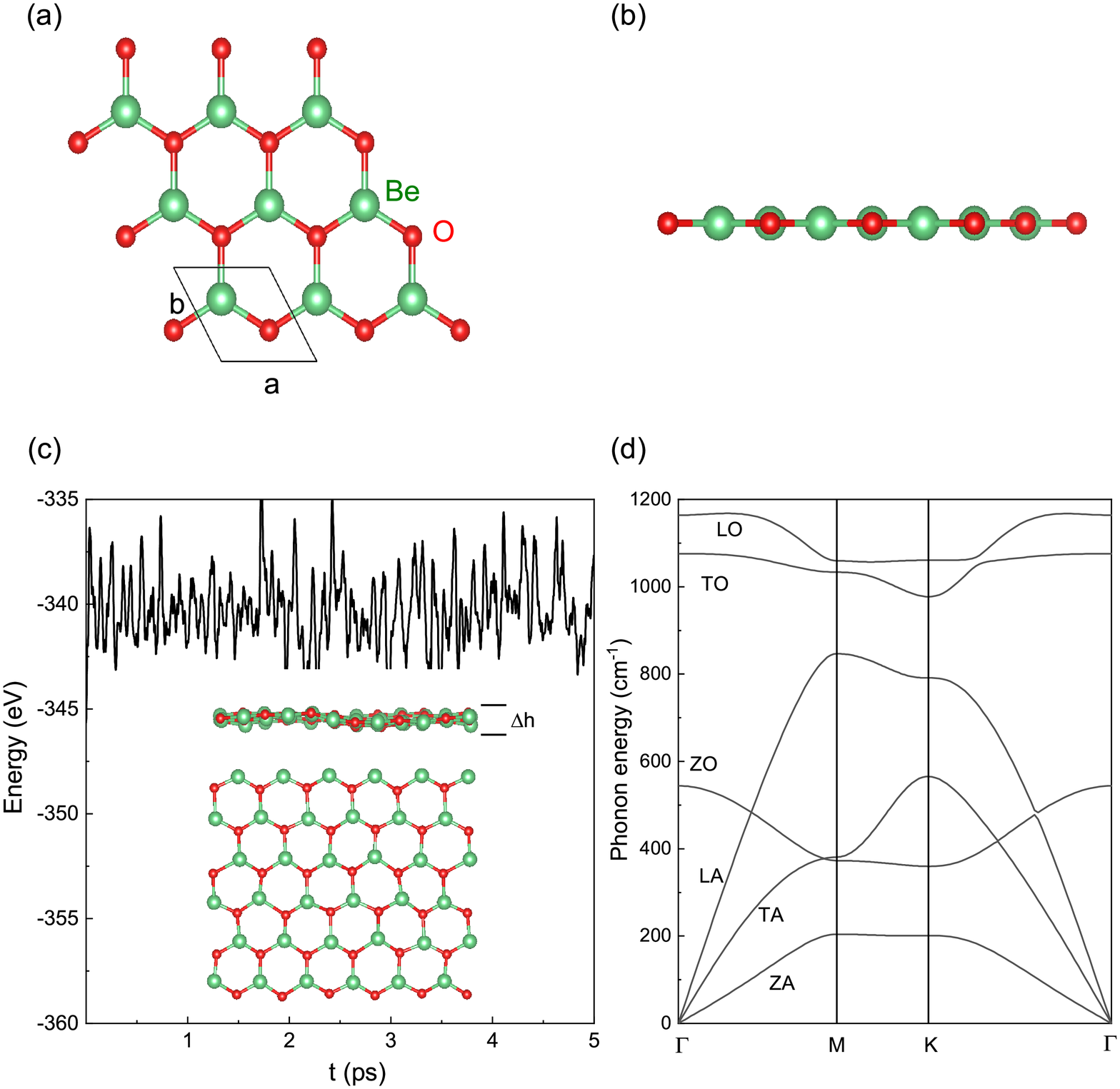}
\caption{\label{fig:structure} (a) Top view and (b) side view of monolayer h-BeO. (c) MD simulation with 800 K and the insets are crystal structure at 5 ps. (d) Phonon dispersions of three acoustic modes and three optical modes.}
\end{figure}

Figure.\ref{fig:structure} shows that the crystal structure of monolayer h-BeO has lattice constants a=b=2.68 \AA\ and is similar to the monoatomic-layer graphene and h-BN. In the calculation of bilayer structure with van der Waals corrected exchange-correlation functionals, AB-type stacking (the atom in upper layer locates the center of the hexatomic ring of the lower layer) has the lowest total energy in the all bilayer structures (see the supplementary material) and the interlayer distance is 3.05 \AA\ , identified as the effective thickness of the monolayer h-BeO. Then the elastic properties, molecular dynamics (MD) simulation, and phonon dispersions are calculated in order to ensure the stability of monolayer h-BeO. Firstly, the stiffness tensor elements of C$_{11}$, C$_{12}$, and C$_{66}$ are 149.13, 36.23, and 58.48 N/m, respectively, also fulfill requirements of Born criterion. Young's module and Poisson's Ratio are evaluated with a value of 162.32 N/m and 0.24, smaller than those of graphene and h-BN. Secondly, the long calculations of MD simulation with two temperatures of 300 K (see the supplementary material) and 800 K show that the total energy oscillates around the stable value and the hexatomic ring is almost undamaged although the atomic fluctuation emerges with $\Delta$h= 0.2 (0.8) \AA\ under 300 (800) K [Fig.\ref{fig:structure}(c) and supplementary material]. Finally, in the phonon dispersion, the absence of imaginary frequency in the entire Brillouin zone shows the dynamic stability. And the lack of other atomic-layer restrictions for the out-of-plane vibration also leads to the significantly lower phonon frequencies of out-of-plane acoustic (ZA) and optic (ZO) modes than those of in-plane acoustic (transverse TA, longitudinal LA) and optical modes (transverse TO, longitudinal LO). More importantly, the polarity of monolayer h-BeO gives rise to the evident LO-TO splitting at $\Gamma$ point with the value of 88 cm$^{-1}$.

\begin{figure}[htp!]
\includegraphics[width=0.5\textwidth]{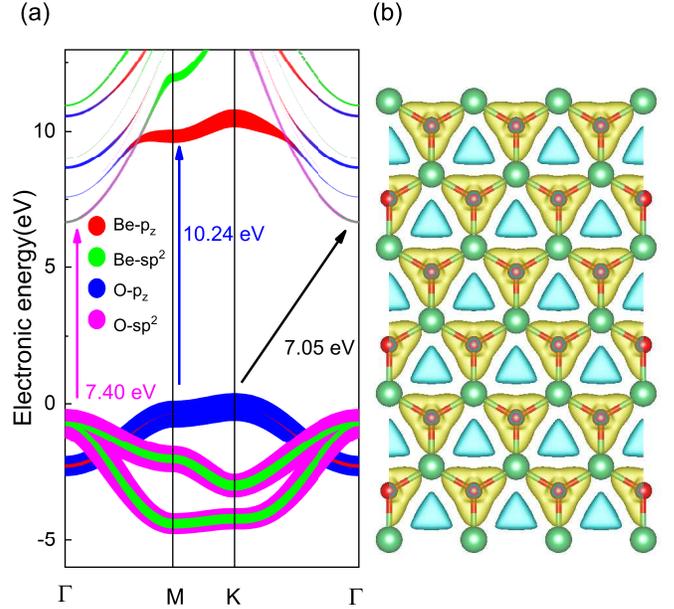}
\caption{\label{fig:band} (a) Calculated electronic structure of monolayer h-BeO with the projections of p$_z$ and sp$^2$ orbits from two elements. (b) Difference charge density $\Delta \rho$. The glassy yellow means increasing charge and light blue means decreasing charge.}
\end{figure}
\begin{table}[htp!]
\caption{\label{tab:table1} The energy gap (E$_{\rm gap}$) between CBM and VBM, the energy gap (E$_{\sigma}$) between bonding $\sigma$ and antibonding $\sigma$*, the energy gap (E$_{\pi}$) between bonding $\pi$ and antibonding $\pi$*, and the effective mass of electron (m*) under different strains with the HSE functional, when the values in the brackets is the results of PBE functional.}
\begin{ruledtabular}
\begin{tabular}{ccccc}
Strain & E$_{gap}$ & E$_{\sigma}$ & E$_{\pi}$ & m*\\
\hline
0  & 7.05 (5.38)  & 7.40 (5.83) & 10.24 (8.95) & 0.792 (0.788)  \\
2  & 6.70   & 7.29 & 9.96  & 0.729 (0.725)  \\
4  & 6.51   & 7.13 & 9.74  & 0.673 (0.680)  \\
6  & 6.28   & 6.95 & 9.52  & 0.641 (0.638)   \\
8  & 6.05   & 6.74 & 9.35  & 0.620 (0.617)   \\
10 & 5.86   & 6.54 & 9.12  & 0.589 (0.590)   \\
\end{tabular}
\end{ruledtabular}
\end{table}
The band structure and charge transfer from Be to O of monolayer h-BeO are presented in Fig.\ref{fig:band}. In the electronic structure, the sp$^2$-hybridization orbitals form the strong in-pane linetype bond, which contribute the bonding $\sigma$ and antibonding $\sigma$* at $\Gamma$ point [Fig.\ref{fig:band}(a)]. And the vertical plane $\pi$ bond is from the p$_z$ orbits.
The electronegative difference between Be and O atoms results in the charge transfer, as the high density around O atoms shown in the contour plots of the difference charge density ($\Delta \rho=\rho_{BeO}-\rho_{Be}-\rho_{O}$) [Fig.\ref{fig:band}(b)]. In the Bader analysis, Be and O have 0.3 and 7.7 charges, respectively. In the classic sp$^2$-hybridization graphene, the purely covalent bond leads to the crossing of $\pi$ and $\pi$* at K point.
But the strong bond in the plane and the high polarization of monolayer h-BeO, from the asymmetry sublattices of Be and O atoms, open an ultrawide bandgap.
The energy gap (E$_{\rm gap}$) between conduction band bottom (CBM) and valence band maximum (VBM), the energy gap (E$_{\sigma}$) between bonding $\sigma$ and antibonding $\sigma$*, the energy gap (E$_{\pi}$) between bonding $\pi$ and antibonding $\pi$* are 7.05, 7.40 and 10.24 eV, respectively, in the HSE functional, which obtain the bigger values than PBE functional [Tab.\ref{tab:table1} and supplementary material].
\begin{figure}[htp!]
\includegraphics[width=0.4\textwidth]{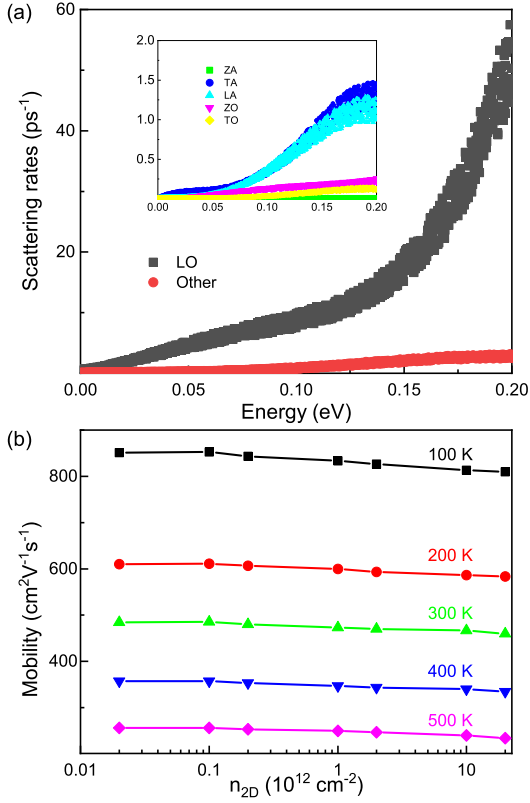}
\caption{\label{fig:mobility} (a) The electronic scattering rates of the LO mode and other modes above CBM at 300 K. The energy of CBM is set to be zero. The inset shows the electronic scattering rates of ZA, TA, LA, ZO, TO modes. (b) The electron carrier mobilities vary with the carrier concentration at different temperatures.}
\end{figure}

The band structure of CBM around $\Gamma$ point is almost quadratic [Fig.\ref{fig:band}(a)] and can be described by the electronic effective mass simply. The $\Gamma$ valley (the band around CBM at $\Gamma$ point) shows an isotropic effective mass m* of 0.792 m$_e$, much lighter than hole effective mass of 3.515 m$_e$, which implies high-performance transport of electronic carrier. In order to reveal the internal electronic transport mechanism, the electronic scattering rates originating from the different phonon modes around CBM are plotted in Fig.\ref{fig:mobility}(a). All the electronic scattering processes in h-BeO are intravalley due to the only $\Gamma$ valley around the edge of the conduction band. As shown, the LO phonons contribute the main electronic scattering rates obviously, which are at least an order of magnitude greater than those of other phonons. The major reason is that the nonzero Born effective charge tensors exist in the polar h-BeO. Then LO phonons can generate the macroscopic polarization field and couple to electrons by the Fr\"{o}hlich interaction~\cite{Verdi2015}.
Moreover, the ZA phonons have little impact on electronic transport, protected by the mirror symmetry~\cite{Fischetti2016}, similar to the graphene.
The calculated electronic carrier mobilities $\mu$ as a function of the carrier concentration at 100-500 K are plotted in Fig.\ref{fig:mobility}(b). The mobility $\mu$ drops noticeably with the increasing temperature at the same carrier concentration because the positive correlation relationship between phonon concentration and temperature causes the strong electron-phonon coupling with increasing temperature.
At room temperature, the mobility $\mu$ has a value of 473 cm$^2$V$^{-1}$s$^{-1}$ with the carrier concentration n$_{\rm 2D}$ =1 $\times$ 10$^{12}$ cm$^{-2}$, when it is as high as 833 cm$^2$V$^{-1}$s$^{-1}$ at 100 K. The slight decrease of $\mu$ with increasing carrier concentration derives from the higher density of electronic states the stronger electron-phonon coupling in $\Gamma$ valley.

\begin{figure}[htp!]
\includegraphics[width=0.4\textwidth]{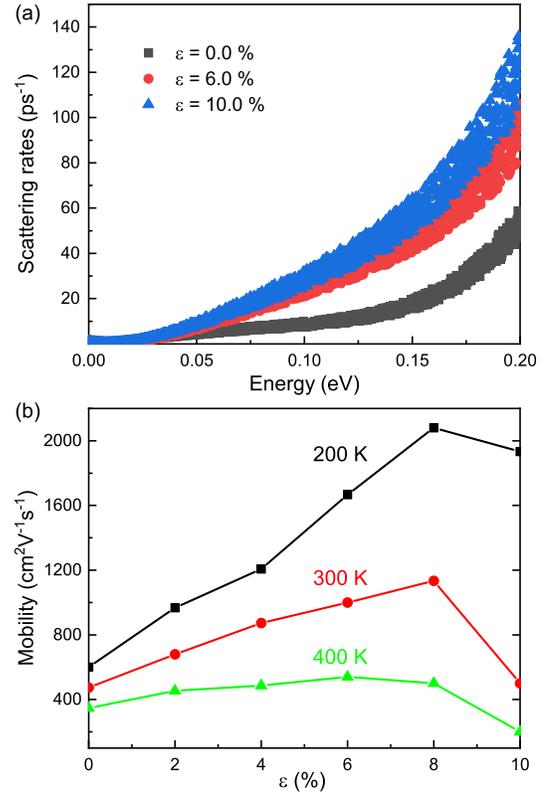}
\caption{\label{fig:strain} (a) The electronic scattering rates of the LO mode above CBM at 300 K with three strains ($\varepsilon$=0.0\%, 6.0\% and 10.0 \%). (b) The electron carrier mobilities vary with the strains at different temperatures when the carrier concentration at 1 $\times$ 10$^{12}$ cm$^{-2}$.}
\end{figure}
Furthermore, the biaxial strain is also calculated due to the distinctive modulation effect on the many physical properties of 2D materials~\cite{Ahn2017,Deng2018} and the compressive strain is out of consideration because of the presence of imaginary frequency around $\Gamma$ point in the phonon dispersions (see the supplementary material). Under the tensile strain, the increasing atomic distances result in the narrowing energy gap between bonding and antibonding states, as the value changes of E$_{\rm gap}$, E$_{\sigma}$ and E$_{\pi}$ summarised in Tab.\ref{tab:table1}. Meanwhile, the electronic effective mass gets lighter under larger strain and reduce to 0.589 m$_e$ with $\varepsilon$=10.0\%. But, it should be also noted that the strain effect on the electron-phonon coupling. Normally, the tensile strain can soften the phonons. And the electron-phonon coupling matrix element are generally inverse correlation with phonon frequency. Hence, there are higher electron-phonon coupling strength and electronic scattering rates under larger strain, as shown in Fig.\ref{fig:strain}(a).
Simultaneously considering various factors of electronic transport, the positive influence of electronic effective mass change, and the negative influence of electron-phonon scattering enhancement, the result is that $\mu$ increases with increasing the degree of strain, and reaches the maximum at $\varepsilon$=8.0 \% before it starts decreasing. At room temperature, suitable strain can raise the mobility to $\sim$1000 cm$^2$V$^{-1}$s$^{-1}$ with n$_{\rm 2D}$ =1 $\times$ 10$^{12}$ cm$^{-2}$, when the mobility can even go up to $\sim$2000 cm$^2$V$^{-1}$s$^{-1}$ at slight lower temperature of 200 K.

In summary, we have systematically investigated the electronic structure, phonon dispersions, electron-phonon coupling, and carrier mobility of monolayer h-BeO using first-principles calculations with Boltzmann transport theory. It is found that monolayer h-BeO is a UWBG material with the indirect bandgap of 7.05 eV in the HSE functional, which is from the two atomic electronegativity difference. The polarity also results in that the LO phonons generate the macroscopic polarization field and couple to electrons by the Fr\"{o}hlich interaction. And the mirror symmetry leads to the negligible effect of ZA phonons on electronic transport. By considering the electron-phonon scattering, monolayer h-BeO has a high mobility of 473 cm$^2$V$^{-1}$s$^{-1}$ at room temperature. In addition, the biaxial tensile strain can reduce the electronic effective mass and enhance the electron-phonon coupling strength. The suitable strain can promote the mobility to $\sim$1000 cm$^2$V$^{-1}$s$^{-1}$ at room temperature.

furthermore, the previous works also find that the many-body effect plays a considerable role in the estimates of bandgap in 2D materials because of the enhanced Coulomb interaction with respect to 3D materials~\cite{Zeng2020,Prete2017,Olsen2016,Latini2017}. We also calculated the quasi-particle bandgap of monolayer h-BeO within the framework of GW approximation~\cite{Huser2013}. The single-shot G$_0$W$_0$ and partially self-consistent GW$_0$ simulations both give the value of bandgap as 8.35 eV.
The results indicate the ultrawide bandgap in monolayer h-BeO is robust to various exchange-correlation functionals and approximate treatments in first-principles calculations.

\begin{acknowledgments}
This work was supported by the National Natural Science Foundation of China (Grants No. 11904312 and No. 11904313),
the Project of the Hebei Education Department, China (Grants No. ZD2018015 and No. QN2018012), and the Natural Science
Foundation of Hebei Province (Grant No. A2019203507). We thank the High Performance Computing Center of Yanshan
University.
\end{acknowledgments}


\begin{thebibliography}{100}
\expandafter\ifx\csname url\endcsname\relax
  \def\url#1{\texttt{#1}}\fi
\expandafter\ifx\csname urlprefix\endcsname\relax\def\urlprefix{URL }\fi
\providecommand{\bibinfo}[2]{#2}


\bibitem{Tsao2018}J. Y. Tsao, S. Chowdhury, M. A. Hollis, D. Jena, N. M. Johnson, K. A. Jones, R. J. Kaplar, S. Rajan, C. G. Van de Walle, E. Bellotti, C. L. Chua, R. Collazo, M. E. Coltrin, J. A. Cooper, K. R. Evans, S. Graham, T. A. Grotjohn, E. R. Heller, M. Higashiwaki, M. S. Islam, P. W. Juodawlkis, M. A. Khan, A. D. Koehler, J. H. Leach, U. K. Mishra, R. J. Nemanich, R. C. N. Pilawa-Podgurski, J. B. Shealy, Z. Sitar, M. J. Tadjer, A. F. Witulski, M. Wraback, and J. A. Simmons, Adv. Electron. Mater. 4, 1600501 (2018).
\bibitem{Xue2018}H. Xue, Q. He, G. Jian, S. Long, T. Pang, and M. Liu, Nanoscale Res Lett 13, 290 (2018).
\bibitem{Wang2020}P. Wang, A. Pandey, J. Gim, W. J. Shin, E. T. Reid, D. A. Laleyan, Y. Sun, D. Zhang, Z. Liu, Z. Zhong, R. Hovden, and Z. Mi,  Appl. Phys. Lett. 116, 171905 (2020).
\bibitem{Chikoidze2019}E. Chikoidze, C. Sartel, H. Mohamed, I. Madaci, T. Tchelidze, M. Modreanu, P. Vales-Castro, C. Rubio, C. Arnold, V. Sallet, Y. Dumont, and A. Perez-Tomas, J. Mater. Chem. C 7, 10231 (2019).
\bibitem{Chae2019}S. Chae, J. Lee, K. A. Mengle, J. T. Heron, and E. Kioupakis, Appl. Phys. Lett. 114, 102104 (2019).
\bibitem{Kim2017}J. Kim, T. Sekiya, N. Miyokawa, N. Watanabe, K. Kimoto, K. Ide, Y. Toda, S. Ueda, N. Ohashi, H. Hiramatsu, H. Hosono, and T. Kamiya, NPG Asia Materials 9, 359 (2017).


\bibitem{Wang2017}J. Wang, F. Ma, W. Liang, R. Wang, and M. Sun, Nanophotonics 6, 943 (2017).
\bibitem{Zhang2017}K. Zhang, Y. Feng, F. Wang, Z. Yang, and J. Wang, J. Mater. Chem. C 5, 11992 (2017).

\bibitem{Song2010}L. Song, L. Ci, H. Lu, P. B. Sorokin, C. Jin, J. Ni, A. G. Kvashnin, D. G. Kvashnin, J. Lou, B. I. Yakobson, and P. M. Ajayan,  Nano Lett. 10, 3209 (2010).


\bibitem{Wang2019}L. Wang, X. Xu, L. Zhang, R. Qiao, M. Wu, Z. Wang, S. Zhang, J. Liang, Z. Zhang, Z. Zhang, W. Chen, X. Xie, J. Zong, Y. Shan, Y. Guo, M. Willinger, H. Wu, Q. Li, W. Wang, P. Gao, S. Wu, Y. Zhang, Y. Jiang, D. Yu, E. Wang, X. Bai, Z. J. Wang, F. Ding, and K. Liu,  Nature 570, 91 (2019).
\bibitem{Liu2020}C. Liu, L. Wang, J. Qi, and K. Liu, Adv. Mater. 32, 2000046 (2020).
\bibitem{Ares2020}P. Ares, T. Cea, M. Holwill, Y. B. Wang, R. Roldan, F. Guinea, D. V. Andreeva, L. Fumagalli, K. S. Novoselov, and C. R. Woods,  Adv. Mater. 32, 1905504 (2020).
\bibitem{Topsakal2009}M. Topsakal, E. Akt\"{u}rk, and S. Ciraci,  Phys. Rev. B 79, 115442 (2009).

\bibitem{Zhao2018}S. Y. F. Zhao, G. A. Elbaz, D. K. Bediako, C. Yu, D. K. Efetov, Y. Guo, J. Ravichandran, K. A. Min, S. Hong, T. Taniguchi, K. Watanabe, L. E. Brus, X. Roy, and P. Kim, Nano Lett. 18, 460 (2018).
\bibitem{Vu2016}Q. A. Vu, Y. S. Shin, Y. R. Kim, V. L. Nguyen, W. T. Kang, H. Kim, D. H. Luong, I. M. Lee, K. Lee, D.-S. Ko, J. Heo, S. Park, Y. H. Lee, and W. J. Yu, Nat. Commun. 7, 12725 (2016).
\bibitem{Wang2015}S. Wang, X. Wang, and J. H. Warner, ACS Nano 9, 5246 (2015).
\bibitem{Cheng2017}R. Cheng, F. Wang, L. Yin, K. Xu, T. A. Shifa, Y. Wen, X. Zhan, J. Li, C. Jiang, Z. Wang, and J. He, Appl. Phys. Lett. 110, 173507 (2017).
\bibitem{Continenza1990}A. Continenza, R. M. Wentzcovitch, and A. J. Freeman, Phys. Rev. B 41, 3540 (1990).
\bibitem{Freeman2006}C. L. Freeman, F. Claeyssens, N. L. Allan, and J. H. Harding, Phys. Rev. Lett. 96, 066102 (2006).

\bibitem{LWang2020}L. Wang, L. Liu, J. Chen, A. Mohsin, J. H. Yum, T. W. Hudnall, C. W. Bielawski, T. Rajh, X. Bai, S. P. Gao, and G. Gu, Angew. Chem. Int. Ed. 59, 2 (2020).


\bibitem{Giannozzi2017}P. Giannozzi, O. Andreussi, T. Brumme, O. Bunau, M. B. Nardelli, M. Calandra, R. Car, C. Cavazzoni, D. Ceresoli, M.
Cococcioni, N. Colonna, I. Carnimeo, A. D. Corso, S. de Gironcoli, P. Delugas, R. DiStasio, A. Ferretti, A. Floris, G. Fratesi, G. Fugallo, R. Gebauer, U. Gerstmann., F. Giustino, T. Gorni, J. Jia, M. Kawamura, H.-Y. Ko, A. Kokalj, E. K\"{u}\c{c}\"{u}kbenli, M. Lazzeri, M. Marsili, N. Marzari, F. Mauri, N. L. Nguyen, H.-V. Nguyen, A. Otero-de-la-Roza, L. Paulatto, S. Ponce, D. Rocca, R. Sabatini, B. Santra, M. Schlipf, A. P. Seitsonen, A. Smogunov, I. Timrov, T. Thonhauser, P. Umari, N. Vast, X. Wu, and S. Baroni, J. Phys.: Condens. Matter 29, 465901 (2017).
\bibitem{Perdew1996}J. P. Perdew, K. Burke, and M. Ernzerhof, Phys. Rev. Lett. 77, 3865 (1996).


\bibitem{Heyd2003}J. Heyd, G. E. Scuseria, and M. Ernzerhof, J. Chem. Phys. 118, 8207 (2003).
\bibitem{Ponce2016}S. Ponce, E. R. Margine, C. Verdi, and F. Giustino, Comput. Phys. Commun. 209, 116 (2016).

\bibitem{Verdi2015}C. Verdi, and F. Giustino, Phys. Rev. Lett. 115, 176401 (2015).
\bibitem{Fischetti2016}M. V. Fischetti and W. G. Vandenberghe, Phys. Rev. B 93, 155413 (2016)
\bibitem{Ahn2017}G. H. Ahn, M. Amani, H. Rasool, D.-H. Lien, J. P. Mastandrea, J. W. Ager III, M. Dubey, D. C. Chrzan, A. M. Minor, and A. Javey, Nat Commun 8, 608 (2017).
\bibitem{Deng2018}S. Deng, A. V. Sumant, and V. Berry, Nano Today 22, 14 (2018).

\bibitem{Prete2017}M. S. Prete, A. M. Conte, P. Gori, F. Bechstedt, and O. Pulci, Appl. Phys. Lett. 110, 012103 (2017).
\bibitem{Olsen2016}T. Olsen, S. Latini, F. Rasmussen, and K. S. Thygesen, Phys. Rev. Lett. 116, 056401 (2016).

\bibitem{Latini2017}S. Latini, K. T. Winther, T. Olsen, and K. S. Thygesen, Nano Lett. 17, 938 (2017).


\bibitem{Zeng2020}M. Zeng, J. Liu, L. Zhou, R. G. Mendes, Y. Dong, M. Y. Zhang, Z. H. Cui, Z. Cai, Z. Zhang, D. Zhu, T. Yang, X. Li, J. Wang, L. Zhao, G. Chen, H. Jiang, M. H. Rummeli, H. Zhou, and L. Fu, Nat. Mater. 19, 528 (2020).




\bibitem{Huser2013}F. H\"{u}ser, T. Olsen, and K. S. Thygesen, Phys. Rev. B 87, 235132 (2013).


\end{thebibliography}
\end{document}